\documentclass{elsart}

\usepackage{amssymb,amsmath}
\usepackage[dvips]{graphicx}
\usepackage{subfigure}

\usepackage{dcolumn,epsfig}

\begin{document}
\begin{frontmatter}

\title{Deterministic inhomogeneous inertia ratchets}

\author{Shantu Saikia}{$^{1,2}$}, and %
\author{Mangal C. Mahato{$^2$}\corauthref{cor}}%
\corauth[cor]{Corresponding Author}%
\ead{mangal@nehu.ac.in}%
\address{{$^1$}St.Anthony's College, Shillong-793001, India}
\address{{$^2$}Department of Physics, North-Eastern Hill University, 
Shillong-793022, India}

\begin{abstract}
We study the deterministic dynamics of a periodically driven particle in the 
underdamped case in a spatially symmetric periodic potential. The system is 
subjected to a space-dependent friction coefficient, which is similarly 
periodic as the potential but with a phase difference. We observe that 
frictional inhomogeneity in a symmetric periodic potential mimics most of the 
qualitative features of deterministic dynamics in a homogeneous system with 
an asymmetric periodic potential. We point out the need of averaging over the 
initial phase of the external drive at small frictional inhomogeneity parameter 
values or analogously low potential asymmetry regimes in 
obtaining ratchet current. We also show that at low amplitudes of the drive, 
where ratchet current is not possible in the deterministic case, noise plays a 
significant role in realizing ratchet current. 

\end{abstract}
\begin{keyword}
Inhomogeneous systems, underdamped systems, deterministic ratchets,
 Langevin equation
\PACS 05.10.Gg, 05.40.-a, 05.40.Jc, 05.60.Cd

\end{keyword}

\end{frontmatter}

\section{Introduction}
Directed particle transport in asymmetric periodic (ratchetlike) potentials without 
the application of any obvious external bias but aided by thermal fluctuations
has been widely studied both in overdamped and underdamped conditions\cite{Reimann,
Astumian,Feynmann,Machura,Ajdari}. This effect, called the ratchet effect, has also 
been obtained in underdamped symmetric periodic (non-ratchetlike) potential systems, 
by considering either the temperature\cite{Landaurer,Buttiker,Blanter,Benjamin} or 
the friction coefficient\cite{Saikia,Reenbohn} to be space dependent. Ratchet 
current can also result in a symmetric periodic potential when the system is driven 
by an externally applied zero-mean temporally asymmetric forcing aided by Gaussian 
white noise\cite{Mahato} or by the explicit application of nonthermal 
noise\cite{Hanggi,Flach}. The ratchet effect is being studied intensively for about 
two decades and its applications envisaged in periodic systems that are common in 
many areas of natural sciences\cite{Hanggi09}.

\par All the above ratchet models use the presence of (Gaussian or colored) noise 
(fluctuations) to obtain directed transport. In these models noise is essential.
However, the study of deterministic ratchets, pioneered by Ref.\cite{Jung}, has 
also contributed significantly to the understanding of the subject. These 
deterministic ratchets, unaided by noise, are shown to yield 
current in overdamped\cite{Bartussek,Chauwin}, underdamped\cite{Flach,Jung,
Kenfack,Mateos1,Barbi1,Arizmendi,Borromeo,Larrondo1,Feng,Marchesoni,Carusela}, 
as well as in Hamiltonian\cite{Oleg,Soskin,Hennig} periodic potential systems, 
and also in overdamped quenched disordered
\cite{Popescu,Family} systems. In these systems net current results, without 
the presence of applied nonzero 
average forcing or asymmetric fluctuations, due to the presence of various 
regular transporting or chaotic attractors depending on the initial conditions for 
given system parameter values. In Ref.\cite{Flach} the symmetry criteria for 
the realization of ratchet current have been discussed in detail.

\par 
Owing to the complexity of the dynamics, for example the simultaneous presence 
of periodic and chaotic attractors, there is a need for proper choice of initial 
conditions and ensemble averaging over them to obtain realistic averages\cite{Flach,
Kenfack,Oleg}. This fact has been exemplified in an (zero mean) ac modulated
periodic Hamiltonian system\cite{Hennig}. It has been shown there that slower 
modulations of the symmetric periodic potential lead to asymmetric access (in
momentum) of chaotic regimes in the phase space resulting in a giant net particle
transport in one direction. The direction and magnitude of the current depends 
on the initial phase of the modulating ac drive. Of course, the net current must 
disappear if averaged over the initial phase.

\par The driven deterministic systems show frequent ensemble-averaged current 
reversals as the drive parameters are changed. Mateos identified bifurcation 
from a chaotic to a periodic regime as the mechanism for the average-current 
reversals\cite{Mateos1} in these systems. However, the change in direction of 
individual single particle trajectories could be related to phase locking 
phenomena\cite{Barbi1} due to the presence of various velocity 
attractors\cite{Larrondo1}. The average current directions are thus sensitive to
initial conditions\cite{Mateos2} which allow to explore these attractors 
selectively in the phase space. In a recent further investigation\cite{Kenfack} 
it has been conjectured that, in general, bifurcations from chaos to periodic 
behaviour could be associated with abrupt changes (and not necessarily reversals) 
in the ensemble averaged currents. 

\par In Ref.\cite{Family}, the effect of different control parameters were 
analysed in the inertial limit. It was found that the control mechanisms were 
associated with the fractal nature of the basins of attraction of the mean 
velocity attractors and that small perturbations of the control variable could 
produce drift reversals. However, the effect of the presence of a weak subharmonic 
component in the ac drive field on the phase locked dynamics of a ratchet is to 
suppress chaos and stabilize regular orbits over large range of driving 
amplitudes\cite{Barbi2}. 

\par The transport properties of inertial systems depend sensitively on the 
(constant) friction parameter\cite{Risken,Marchesoni,Carusela}. The addition of 
noise to the system in the low damping regime makes the system dynamics robust 
against initial conditions\cite{Marchesoni}. Moreover, in a well defined damping 
window the efficiency of ratchet current generation becomes appreciable even for 
small applied periodic field amplitudes. In Ref. \cite{Carusela} the effect of 
noise on individual transporting deterministic trajectories is analysed. A small 
noise seem to have little effect on some trajectories in one direction whereas it 
destroys trajectories in the opposite direction resulting in a robust finite 
ratchet current.
 
\par In most of the above works (barring a few, for instance\cite{Flach,Oleg,
Hennig}), particle motion in periodic but asymmetric potentials is considered; 
for example a system with potential $V(x)=V_0(\cos x + b\cos 2x)$ driven by a
 periodic field, $F(t)=a\cos(\omega t +\phi_0)$ in uniform friction media. 
In these studies the potential asymmetry ($b\neq 0$) is the primary cause of net
 particle drift. In the present work, we consider a {\it{symmetric}} periodic 
potential, for example
$V(x)=-V_0\sin x$, but in the presence of a space dependent friction coefficient
$\gamma(x)=\gamma_0(1-\lambda\sin(x+\phi))$ (as in \cite{Saikia}).
A massive charged particle moving in a periodic ionic lattice and in a medium
with material density (friction) profile created by a stationary pressure wave
can be thought of as a mechanical illustration of such
systems\cite{Saikia,Reenbohn}. However, such systems have been studied earlier
in a resistively and capacitatively shunted junction circuit model of small
Josephson junctions\cite{Falco,Barone}. The frictional inhomogeneity term
emulating there the "$\cos\theta$" term representing the coupling between
the quasiparticle tunneling and Cooper pair tunneling across the Josephson
junction. In Ref. \cite{Falco} the phase space trajectories of this system are
presented in detail.

In the present work we show that all the features of deterministic inertia 
ratchets can be observed even in the symmetric periodic potential system in an
 inhomogeneous medium.
The frictional nonuniformity of the medium with phase shift $\phi(\neq 0,\pi)$ 
essentially emulates the effect of the asymmetry of the potential considered in
earlier cases. Moreover, we show in our present inhomogeneous dissipative system 
that, for small inhomogeneity (roughly $\lambda\leq 0.6$), in order to get sensible 
ratchet current one also needs to average over the initial phase $\phi_0$ of the 
drive as in the Hamiltonian system\cite{Oleg,Hennig}. Though ratchet current is a 
steady state ($t\rightarrow \infty$) phenomenon one is likely to obtain a finite 
ratchet current, an obvious erroneous result, even at zero potential asymmetry in a 
uniform medium without this averaging. Interestingly, the bifurcation diagrams with 
and without the frictional inhomogeneity are qualitatively indistinguishable. However,
the stroboscopic Poincar\'{e} plots differ qualitatively in the driven system as the
inhomogeneity parameter $\lambda$ is varied. 
 Links
Vi
The presence of noise (which is ubiquitous in natural environments) smooths out 
the current fluctuations so numerously present in the deterministic 
case \cite{Carusela, Denisov}. In Ref. \cite{Denisov} a matrix continued fraction 
method is presented to obtain ratchet current in noisy periodically driven 
inertial systems. We further show in our inhomogeneous system that at low 
drive amplitudes noise plays a crucial role in getting ratchet (mean) current, 
that is conspicuously absent in the deterministic drive case, though at larger 
drive amplitudes one obtains ratchet current without the presence of noise.  

\section{The model}

As mentioned earlier, in this work, we consider the motion of a particle in a 
periodic potential $V(x)=-V_0 \sin(kx)$ which is symmetric in space (about $kx=
(2n+1)\pi/2$, $n=0, \pm1, \pm2, ...$). The friction coefficient 
$\gamma (x)=\gamma_0(1-\lambda \sin(kx+\phi))$ is periodic with the same 
periodicity as the potential but has a phase difference. The system is driven 
periodically by an external periodic forcing 
$F(t)$=$a$ $\cos(\omega t + \phi_0)$. We study the system in the deterministic 
regime i.e. in the absence of noise and also at the end in the presence of noise. 

The one dimensional equation of motion of a particle of mass $m$ is given by
the Langevin equation,
\begin{equation}
m\frac{d^{2}x}{dt^{2}}=-\gamma (x)\frac{dx}{dt}-\frac{\partial{V(x)}}{\partial
x}+F(t).
\end{equation}
 The corresponding equation of motion in the presence of noise is given by
\begin{equation}
m\frac{d^{2}x}{dt^{2}}=-\gamma (x)\frac{dx}{dt}-\frac{\partial{V(x)}}{\partial
x}+F(t)+\sqrt{\gamma(x)T}\xi(t).
\end{equation}
In Eq.2, $T$ is the temperature in units of the Boltzmann constant $k_B$.
The Gaussian distributed fluctuating forces $\xi (t)$ satisfy the statistics:
 $<\xi (t)>=0$, and $<\xi(t)\xi(t^{'})>=2\delta(t-t^{'})$. 
For convenience, we write down Eq.1 and Eq.2 in 
dimensionless units by setting $m=1$, $V_0=1$, $k=1$. In Eq.2, $T=2$ then corresponds 
to an energy equivalent equal to the potential barrier height at $a=0$. In terms of 
the reduced variables denoted again now by the same symbols, the two equations
can be written as 
\begin{equation}
\frac{d^{2}x}{dt^{2}}=-\gamma(x)\frac{dx}{dt}
+\cos x +F(t),
\end{equation}
and
\begin{equation}
\frac{d^{2}x}{dt^{2}}=-\gamma(x)\frac{dx}{dt}
+\cos x +F(t)+\sqrt{\gamma(x) T}\xi(t),
\end{equation}
where $\gamma(x)=\gamma_0(1-\lambda \sin(x+\phi))$. Thus the periodicity of the 
potential $V(x)$ and also the friction coefficient $\gamma$ is $2\pi$ . The 
potential barrier between any two consecutive wells of $V(x)$ persists for all field
amplitudes $a$, ($0<a<1$) and it just disappears at the critical field
 value $a=a_c=1$.
Notice that for $\phi=n\pi$, $n=0,\pm 1,\pm 2, ....$ the deterministic Eq.3 should 
yield exactly the same trajectory but in the opposite direction on letting 
$x\rightarrow x-\pi/2$ and $t\rightarrow -t$ and in the presence of a thermal noise
Eq.4 should give the same ensemble averaged current but again in the opposite
direction\cite{Flach}. Thus in order that a ratchet current appears the symmetry 
of the system must be broken using $\phi\neq n\pi$.

The noise variable, $\xi$ in Eq.4, satisfies exactly similar statistics as earlier. 
Eq.3 and Eq.4 are solved numerically using the Heun's method for solving 
differential equations. The particle trajectories are obtained for given initial 
conditions for fixed values of $a$ and $\gamma_0$. Also, the steady state mean 
velocity of the particles are calculated.
\par  For the deterministic case the average velocity for a single particle 
(with a given initial condition) is obtained by the relation 
\cite{Larrondo1}
\begin{equation}
 {\langle v \rangle}= \frac{x(t_{max})-x(t_{tran})}{(t_{max}-t_{tran})},
\end{equation}
where $t_{max}$ is the maximum time for which the particle is allowed to evolve and 
$t_{tran}$ is the duration of initial transients that we remove. 
\par Also the mean velocity averaged over an ensemble of $N$ particles 
is defined as 
\begin{equation}
 {\langle \langle v \rangle \rangle}= \frac{1}{N}\sum_{i=1}^{N}{\langle v \rangle}_i,
\end{equation}

For the particle trajectory in the presence of noise, the steady state mean velocity
$\bar v$ of the particle is obtained as
\begin{equation}
 {\bar v}=   \langle \lim_{t \rightarrow \infty}\frac{x(t)}{t}\rangle,
\end{equation}
where the average $\langle ... \rangle$ is evaluated over a large number of 
trajectories.

\section{Numerical Results}

The phase space trajectories of the deterministic Eq.3 was investigated earlier
in Ref. \cite{Falco}. Also, for symmetric systems ($\lambda=0$, in Eq.3) the 
regimes of chaotic (nontransporting) and periodic motion were given in detail in
Ref. \cite{Huberman}. In the present work the particle is allowed to evolve for 
a long time so that the system reaches a steady state condition to investigate 
ratchet effect. We solve Eq.3 (for the deterministic case) and Eq.4 (when noise 
is included) using the second order Heun's method with time step size $\Delta 
t = 0.001$. For the calculation of deterministic average velocity (Eq.5) we take 
$t_{max}=10^6$ and $t_{tran}\approx 10^4$. Throughout our work, unless otherwise 
stated, we keep $\lambda = 0.9$, $\gamma_0 = 0.12$ and $\phi = 0.35$. We partly
repeat the investigations done earlier \cite{Mateos1,Barbi1,Larrondo1} replacing
 the asymmetric periodic potentials by a symmetric periodic potential but instead
 of a uniform-friction-medium we consider here a periodically varying friction 
coefficient and present our results below.

\begin{figure}
\begin{center}
\epsfig{file=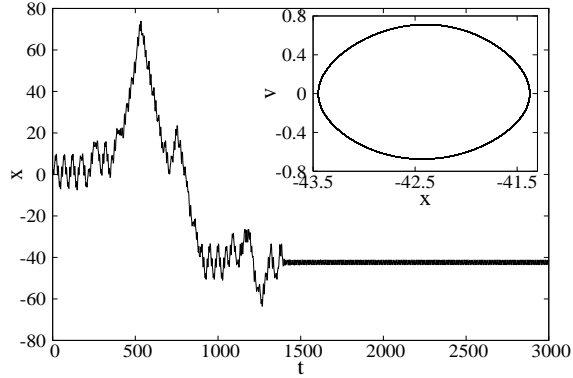, angle=-90, width=7.5cm,totalheight=5cm}
\caption{Plot of particle trajectory for $a = 0.5$ and $\tau = 10$ (main figure) and the 
corresponding phase space trajectory; $x(0)=\pi/2$, $v(0)=0$, $\gamma_0 = 0.12$, 
$\phi = 0.35$, $\lambda = 0.9$.}
\end{center}
\end{figure}

\begin{figure}
\begin{center}
\epsfig{file=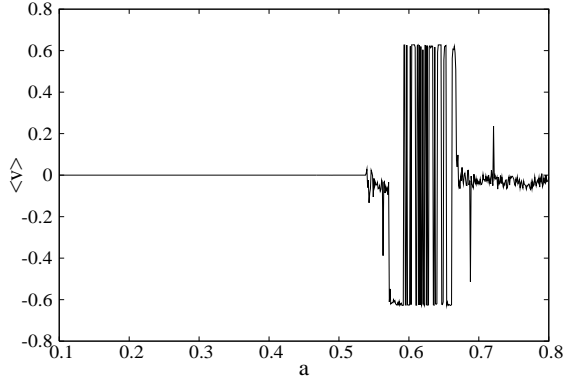, angle=-90, width=7.5cm,totalheight=5cm}
\caption{Variation of mean velocity $\langle v \rangle$ with amplitude $a$ for a single
particle; $v(0) = 0$, $x(0)$ = $\pi/2$, $\gamma_0 = 0.12$, $\phi = 0.35$,
 $\lambda = 0.9$ .}
\end{center}
\end{figure}

\subsection{Phase locked dynamics and sensitivity to initial conditions}
The dynamics of the particle is critically dependent on the amplitude $a$ 
and frequency $\omega$ of the zero-mean external periodic drive\cite{Family}. 
The dissipative system goes to the only fixed point attractor at $v=0, 
x=(2n-1)\pi /2$ when the system is not driven ($a=0$). For a given (non-zero) 
amplitude and frequency of the drive the individual particle trajectories depend 
sensitively on the initial conditions. In the low amplitude regime (below a 
particular critical amplitude depending on the frequency) the particle, after 
initial transients, gets invariably trapped to a periodic attractor in some 
potential minima $x=(2n+1)\pi /2$ ($n=0,\pm1,\pm2,...$) and executes periodic 
motion within the potential well (Fig.1). On the average these cases give zero 
net current as is evident from the phase space trajectory (Inset, Fig.1). 
Only beyond certain amplitude $a$ (for a given initial condition and drive 
frequency $\omega$), the particle starts evolving giving rise to either 
periodic or chaotic dynamics. 

Fig.2 shows the mean velocity $<v>$ for a 
typical single particle trajectory (with the same initial condition ($x(0)=\pi/2$ and
$v(0)=0$ ) as a function of drive amplitude $a$ and 
period $\tau=10$, where $\tau=2\pi/\omega$. There are numerous changes in the 
direction of current as the amplitude $a$ is set at different values. For 
example, whereas $a=0.5931$ gives a positive $<v>$, $a=0.592$ for the same 
initial conditions gives mean velocity -$<v>$. This behaviour can be traced to 
amplitude dependent phase locked periodic behaviour of particle motion as 
shown in Figs.3a,b. As the border between the basins of these attractors are
crossed with the variation of the amplitude of the drive the single particle 
shows a change in the direction of $<v>$ \cite{Barbi1,Larrondo1}.

\begin{figure}[t]
  \centering
  \subfigure[]{\includegraphics[width=5cm,height=6.5cm,angle=-90]{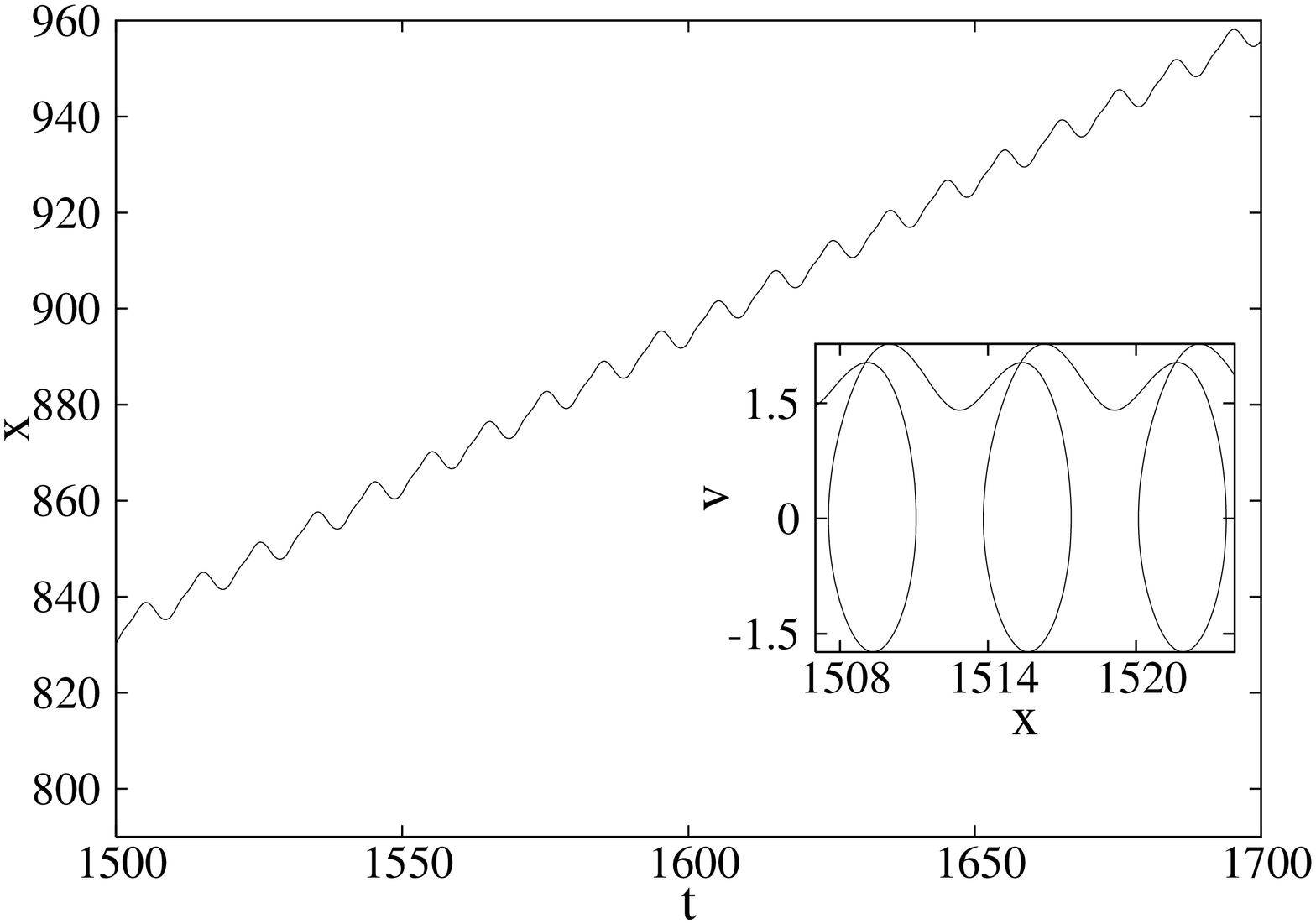}}
\hspace{0.4cm}
\subfigure[]{\label{fig:edge-c}\includegraphics[width=5cm,height=6.5cm,angle=-90]
{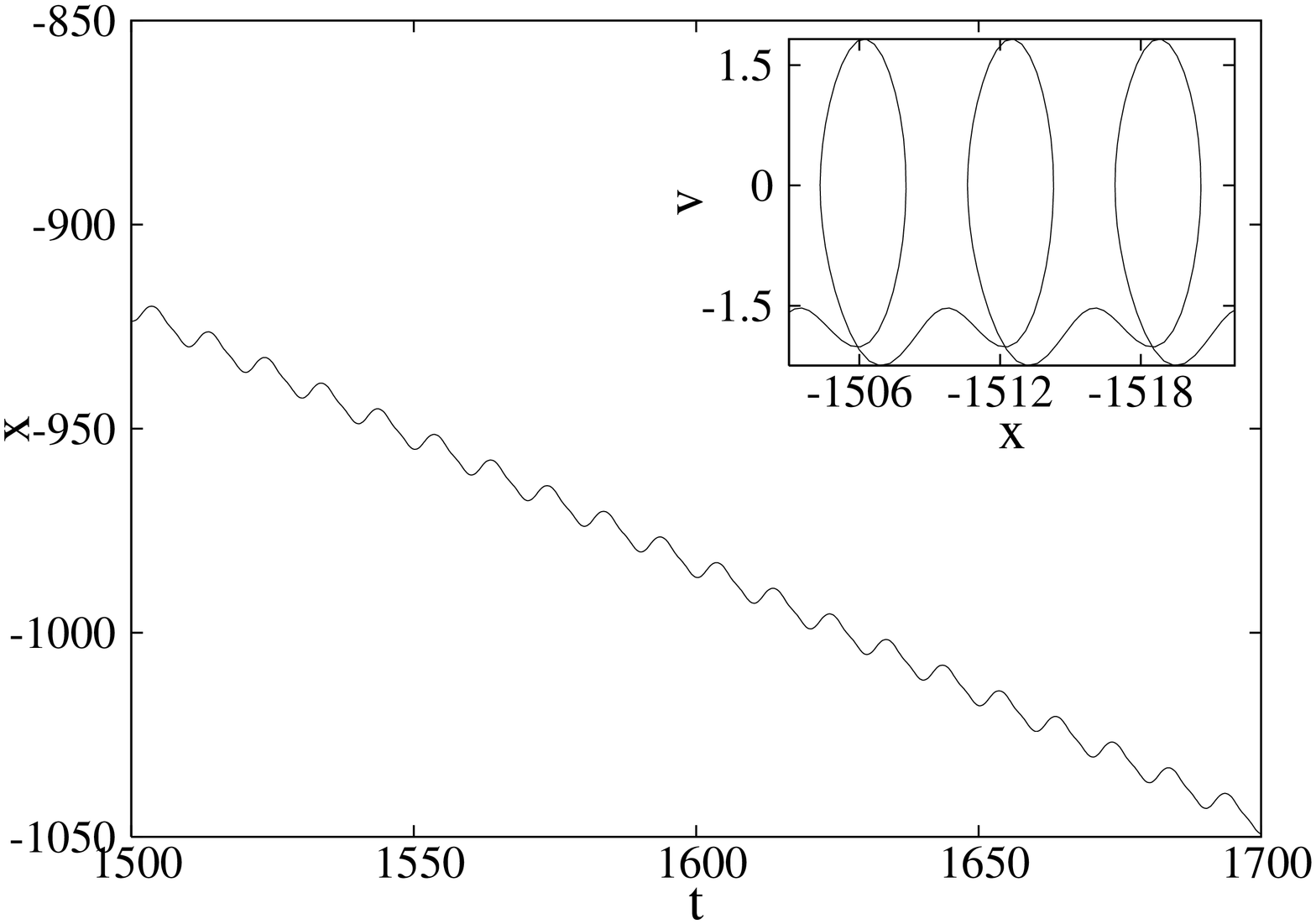}}
\caption{Particle trajectories (main figures) and the corresponding phase space
 plots (insets) showing different locking behaviour of the particle. For $a = 0.5931$ 
(fig. a), the particle gets locked in the  positive sense while for $a = 0.592$ (fig. b), 
the particle gets locked in the negative sense. $\tau = 10$, $x(0) = \pi/2$, $v(0) = 0$
 for both cases.}
\label{fig:edge}
\end{figure}

 Also, the sensitivity of the trajectories to initial 
conditions can be inferred by the following observation. For $a=0.7$ and 
$\tau=10$, the intial condition $v(0)=0, x(0)=1.38309$ gives a transporting 
phase locked trajectory in the positive direction as in Fig. 3a, whereas $v(0)=0, 
x(0)=1.383093$ gives locked periodic nontransporting trajectory as in Fig.1 and
$v(0)=0, x(0)=1.38309704$ results in a transporting phase locked trajectory in 
the negative direction as in Fig. 3b. The sensitivity of the initial conditions
demands that meaningful net particle velocity for a given $a$ and $\omega$ 
be obtained by averaging over a large number of particle trajectories 
corresponding to various initial conditions.

\par As the nature of the particle trajectory depends critically on the chosen 
initial condition, for realistic averages we calculate the velocity (Eq.6), 
averaged over an ensemble of $N$ (typically equal to 500) particles. The particles 
are taken with identical initial velocities ($v_{ini}=0$), but with initial 
positions uniformly distributed between two consecutive potential 
maxima\cite{Mateos1}. The particle trajectories, and also the phase space 
diagrams, are studied to reveal the characteristics of the particle dynamics. 
The bifurcation diagrams are obtained by recording the values of $\dot x(t_p)$ 
at times $t_p=n_p\tau$, for each value of the control parameter (the amplitude of 
the external drive), where $\tau = 2\pi/\omega$ is the period of the external drive 
and $n_p$ is an integer.
 
\subsection{Ratchet current and bifurcation diagrams}

\begin{figure}[ht]
  \centering
  \subfigure[]{\includegraphics[width=5cm,height=6.5cm,angle=-90]{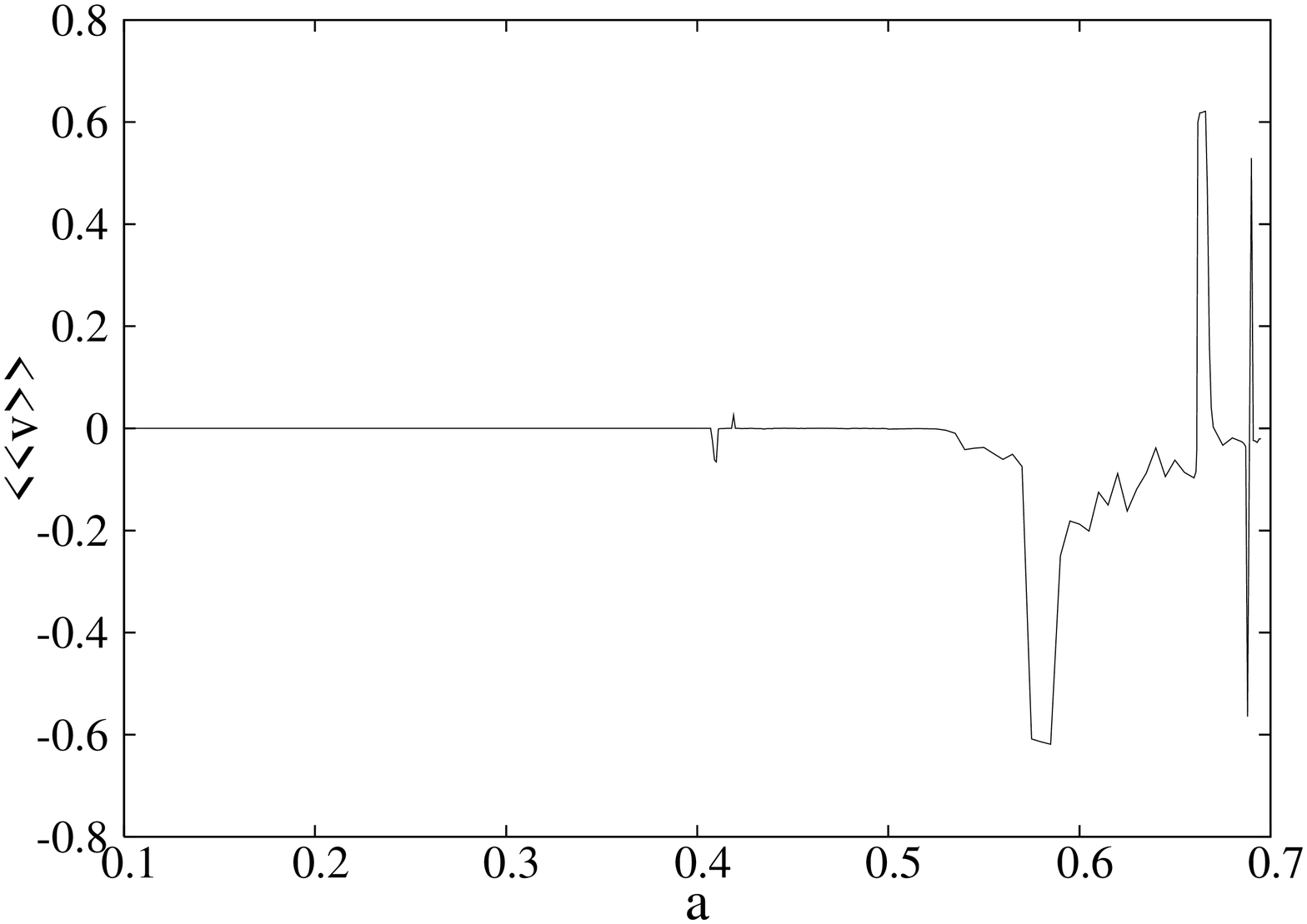}}
\hspace{0.4cm}
\subfigure[]{\label{fig:edge-c}\includegraphics[width=5cm,height=6.5cm,angle=-90]{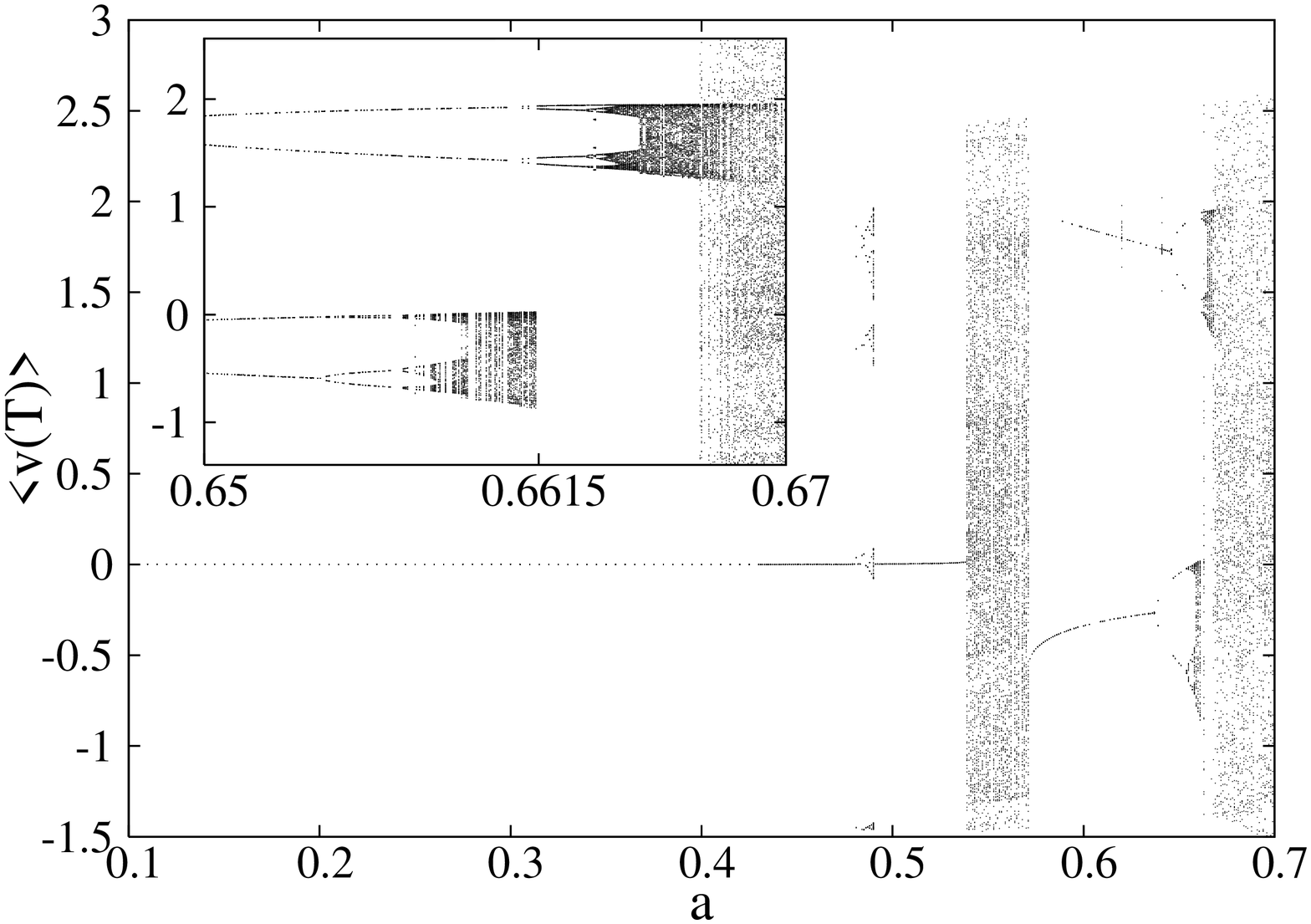}}
\caption{Variation of ensemble averaged velocity $\langle \langle v \rangle \rangle$ 
with amplitude of drive $a$ (a). Fig.4b. shows the corresponding 
bifurcation diagram. Inset of (b) shows the amplitude at which the current reversal occurs
in Fig.4a.  $\tau = 10$, $\gamma = 0.12$, $\lambda = 0.9$, $\phi = 0.35$.}
\label{fig:edge}
\end{figure}

On ensemble averaging, ratchet currents (Fig.4a) are obtained in the deterministic 
inertial inhomogeneous system with a symmetric and periodic potential. The 
averaging is done with the same set of parameters as in Fig.2. The system shows 
current reversals as a function of the amplitude of the drive. However, most of the 
single particle current reversals appearing in Fig.2 disappear. This is due to 
the fact that different initial conditions select different attractors for a 
particular amplitude, all the other parameters remaining the same. So, at a 
particular amplitude of drive, some particles of the ensemble may move in the 
positive direction and some may move in the negative direction with different 
multiples of the fundamental locking velocity $v_{\omega}$ \cite{Barbi1}. Others 
may show a chaotic dynamics giving rise to either zero velocity or some non-zero 
velocity on an average. The corresponding bifurcation diagram (Fig.4b) reflects 
this argument. 

Fig.4b shows that the particle exhibits regimes of periodic and chaotic dynamics 
as a function of the control parameter $a$. The current reversal at $a=0.6615$ in 
Fig.4a is due to a bifurcation from a chaotic to a periodic regime (inset of 
Fig.4b ) which agrees with Mateos conjencture \cite{Mateos1}. However, there 
could be many such bifurcations but without an accompanying current 
reversal\cite{Kenfack}.

\subsection{Frequency dependence of particle dynamics}

\begin{figure}
\begin{center}
  \centering
  \subfigure[]{\label{fig:edge-b}\includegraphics[width=5cm,height=6.5cm,angle=-90]{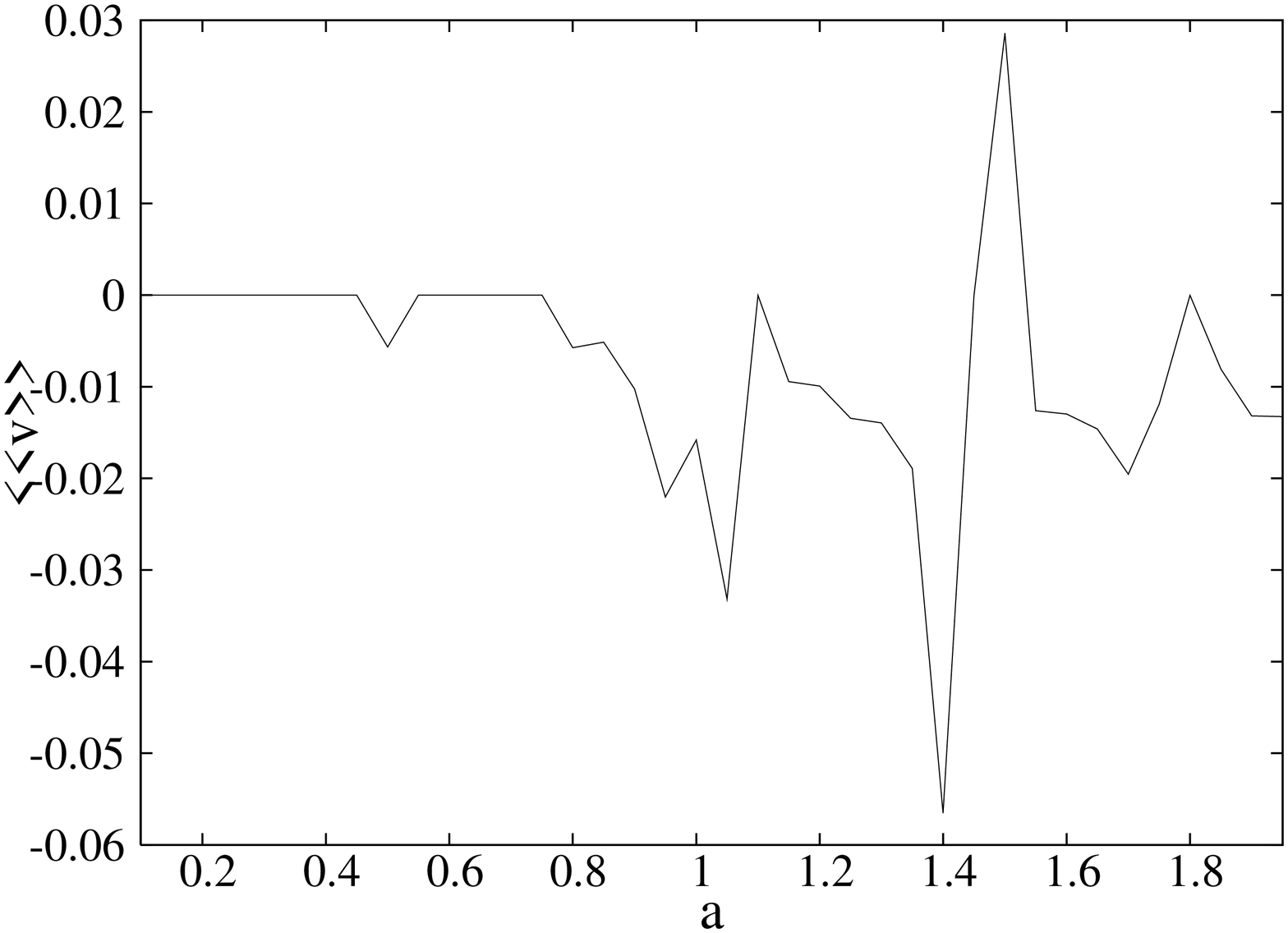}}
\hspace{0.4cm}
\subfigure[]{\label{fig:edge-c}\includegraphics[width=5cm,height=6.5cm,angle=-90]{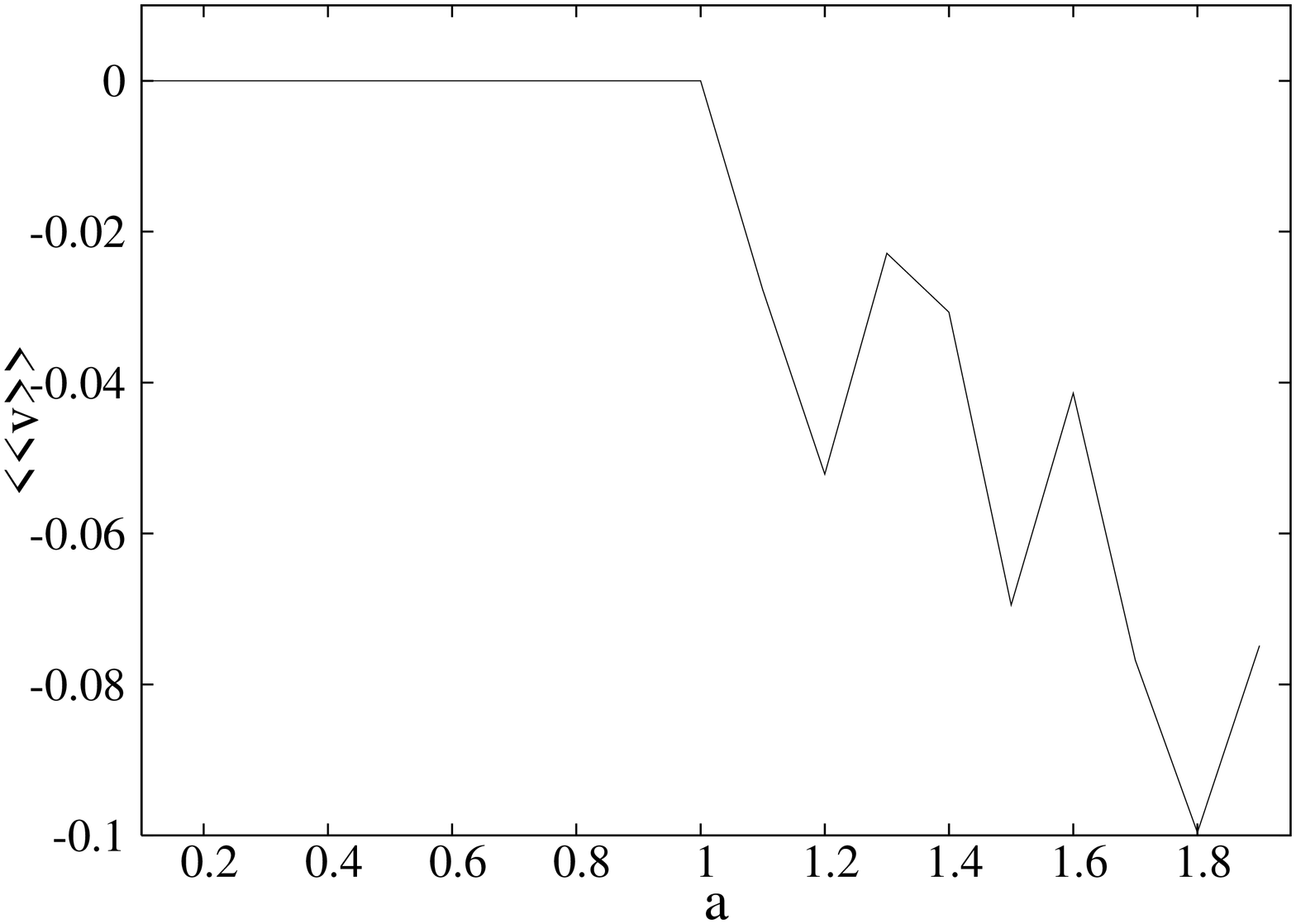}}

  \caption{Variation of ensemble averaged velocity $\langle \langle v \rangle \rangle$ 
with amplitude of drive $a$ for $\tau =100$ (a) and $\tau = 1000$ (b); $\gamma = 0.12$,
 $\lambda = 0.9$, $\phi = 0.35$. }
\label{fig:edge}
\end{center}
\end{figure}

The nature of the particle dynamics i.e. the nature of the current and its 
reversals changes with the driving frequency $\omega=2\pi/\tau$. A comparison 
between the current profile for three periods $\tau = 10$ (Fig.4), $\tau =100$ 
and $\tau= 1000$ (Fig.5) reflects this fact. For different frequencies, particle 
current starts developing at different amplitudes. This is because the amplitude of 
the drive at which the particle starts evolving out of a single-well trapped-state 
differs with frequency. For very low frequencies of the drive $\tau = 1000$ 
(Fig.5b), particle currents are obtained only beyond the critical tilt ($a=1.0$ 
for our case) whereas for $\tau = 10$ current starts developing essentially around
$a=0.55$. Also, whereas for the slower drive $\tau = 1000$ there is no current 
reversal in the range of drive amplitudes shown, for $\tau = 10$ and $100$ there 
are several reversals of the direction of current in the same range of $a$. Thus, 
the mean velocity is expected to show a very complex behaviour in the 
($a-\omega$) space\cite{Family}.

\subsection{Role of frictional inhomogeneity and dependence on initial driving phase}

\begin{figure}[t]
  \centering
  \subfigure[]{\includegraphics[width=5cm,height=6.5cm,angle=-90]{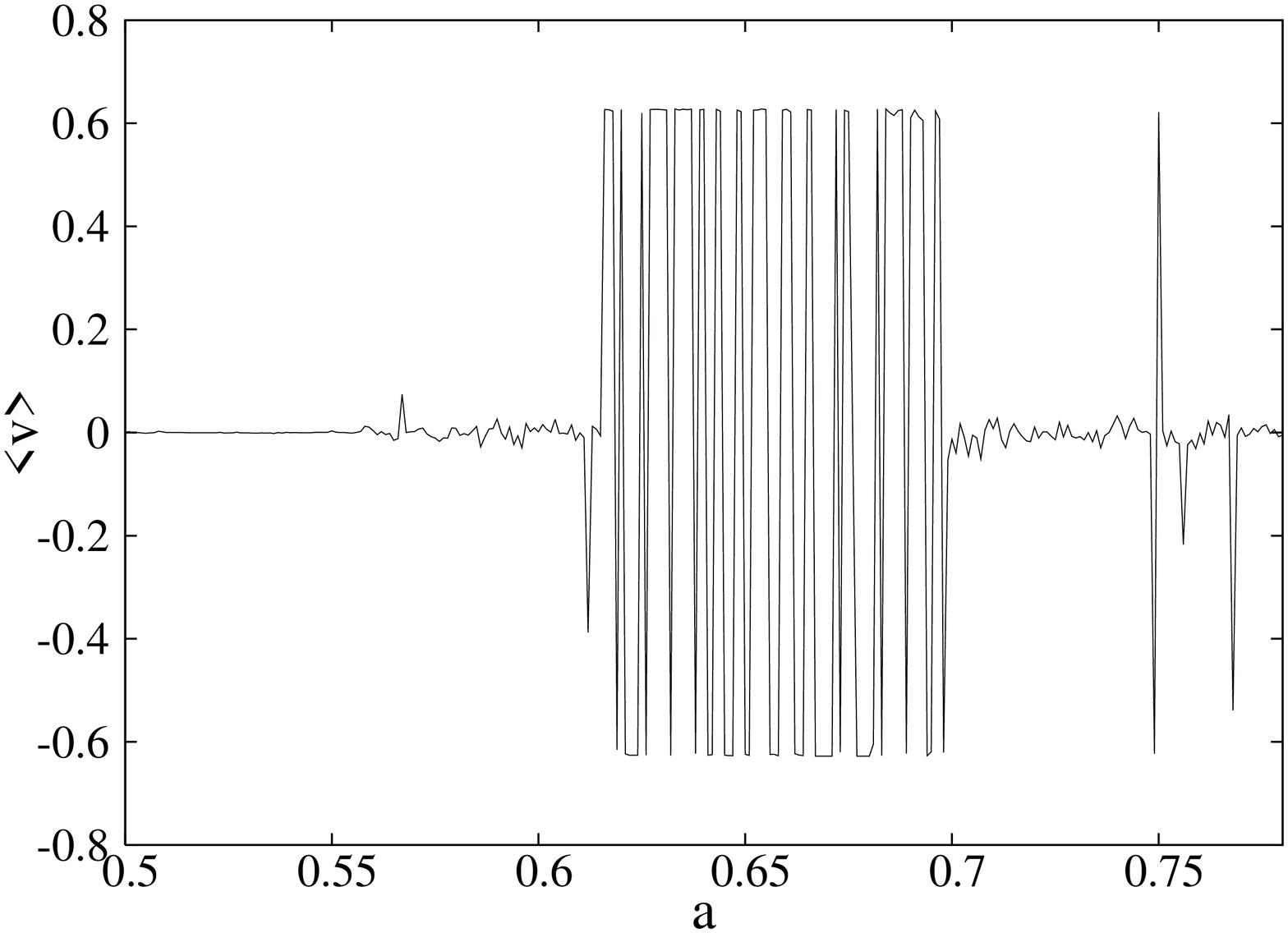}}
\hspace{0.4cm}
\subfigure[]{\label{fig:edge-c}\includegraphics[width=5cm,height=6.5cm,angle=-90]{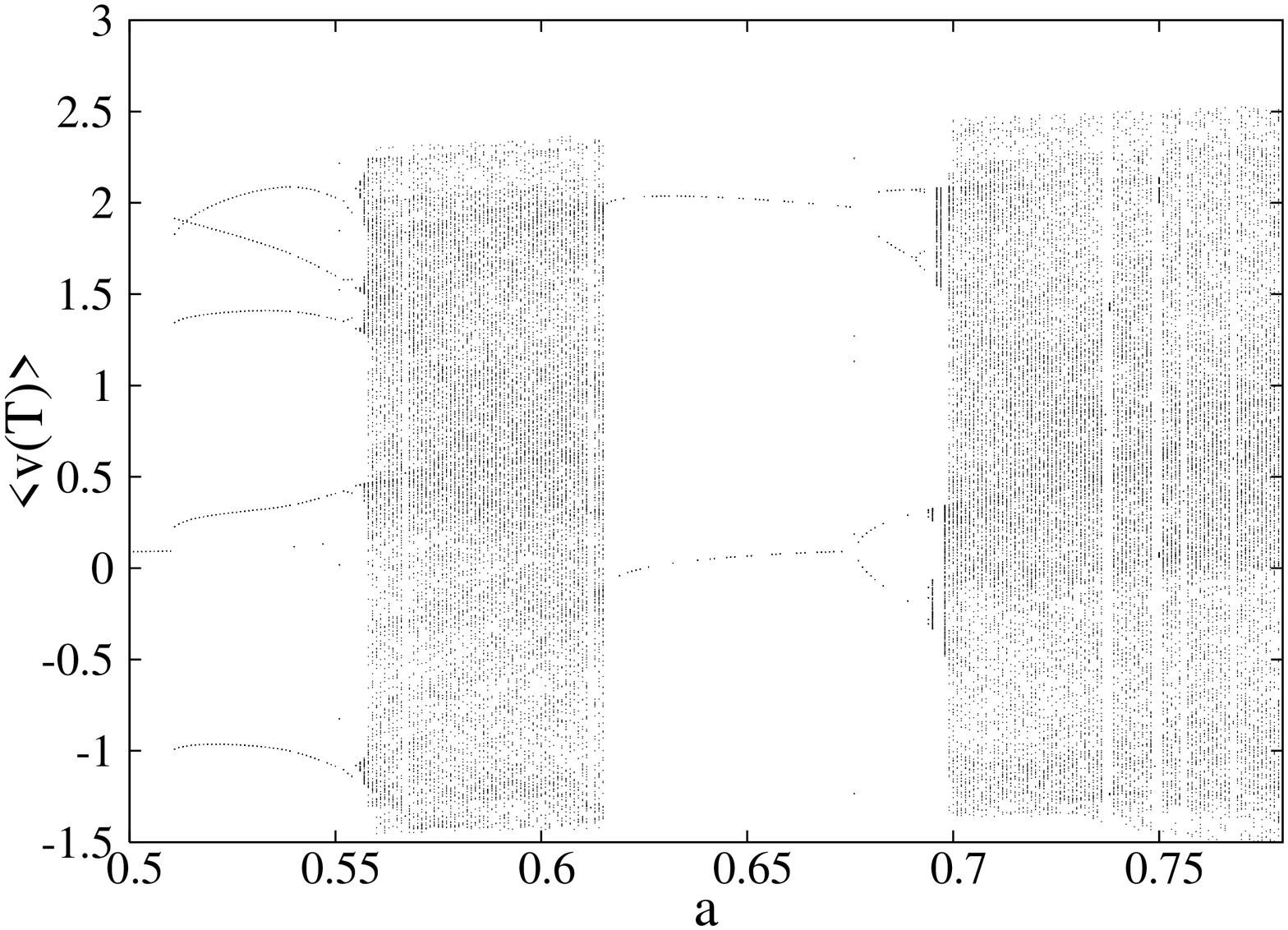}}
\caption{$\langle v \rangle$ versus $a$ plot for a single particle with $\lambda = 0$ (a)
 and the corresponding bifurcation diagram (b). $\tau = 10$, $x(0) = \pi/2$, $v(0) = 0$.}
\label{fig:edge}
\end{figure}

The particle dynamics in the present dissipative system with different degrees 
of frictional inhomogeneity ($\lambda$), for $\phi=0.35$, starting with 
$\lambda=0$ is investigated. We note that $\lambda$ in the present 
work plays a role analogous to the potential asymmetry in the uniform friction 
case. Surprisingly, at certain drive amplitudes,  even with zero inhomogeneity, 
non-zero (steady state, $t\rightarrow \infty$) current is obtained, both at 
the single realisation (Fig.6) case, namely $\langle v\rangle$, as well as when 
averaged over a large number of realizations, $\langle\langle v\rangle\rangle$ 
(Fig.7). Instead, one would have expected the dissipative system to lose the 
initial memory by the time the transients get settled down. The velocities 
$\langle v\rangle$ wildly change with drive amplitude (Fig. 6(a)) but are 
finite. $\langle\langle v \rangle\rangle$ are not only finite but they vary 
smoothly with drive amplitude. This is obviously an erroneous result. Moreover, 
the bifurcation diagrams with inhomogeneity (Fig. 4(b)) and without inhomogeneity
 (Fig.6(b)) on comparison appear very similar and do not provide any clue to the error. 
The error could only come from lack of averaging over the initial phase $\phi_0$ of 
the zero-mean external periodic drive $F(t)$. The relevance of averaging over 
initial phase has earlier been discussed for Hamiltonian 
systems\cite{Oleg,Hennig}.

\begin{figure}
\begin{center}
\epsfig{file=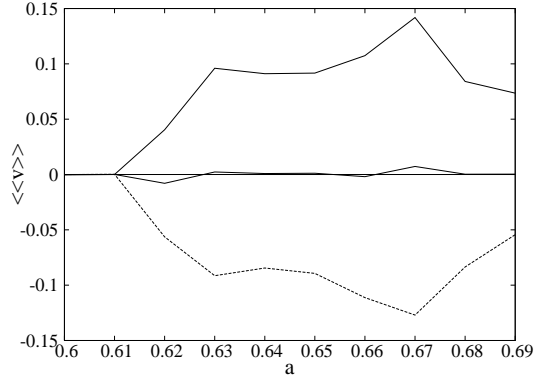, angle=-90, width=7.0cm,totalheight=5cm}
\caption{Plot of $\langle \langle v \rangle \rangle$ versus $a$ with zero inhomogeniety
($\lambda = 0$) for initial phase of the external drive $\phi_0 = 0$ (lower dotted line)
and $\phi_0 = \pi$ (upper solid line). Middle line shows $\langle \langle v \rangle 
\rangle$ after averaging over initial phases. Zeroline is drawn for reference.
 $\tau = 10$, $\lambda = 0.9$, $\gamma = 0.12$, $\phi = 0.35$.}
\end{center}
\end{figure}

In Fig.7, the ensemble averaged current $\langle\langle v \rangle\rangle$ 
are shown for two values of the initial phase $\phi_0=0$ and $\pi$ of the 
external drive $F(t)$ of period $\tau=10$. Clearly, the initial bias due to a 
fixed value of $\phi_0$ determines the steady state current. In order to get a 
sensible ratchet current one needs, therefore, to average over the initial 
phase $\phi_0$ too. The figure also indicates that averaging should be carried 
out over at least two values of $\phi_0$ which differ by $\pi$ .

\begin{figure}
\begin{center}
\epsfig{file=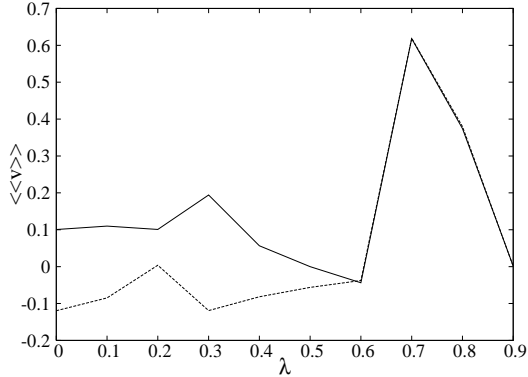, angle=-90, width=7.0cm,totalheight=5cm}
\caption{ Variation of $\langle\langle v \rangle\rangle$ with the asymmetry parameter
$\lambda$ for initial phase of the external drive $\phi_0 = 0$(lower dotted line) and 
$\phi_0 = \pi$(upper solid line); $\tau = 10$, $a = 0.67$, $\gamma = 0.12$, $\phi = 0.35$.}
\end{center}
\end{figure}

 Interestingly, this averaging is required only for small inhomogeneities (Fig.8).
 For large asymmetries $\langle\langle v \rangle\rangle$ does not seem to depend on the 
initial phase $\phi_0$. It turns out that the critical inhomogeneity 
$\lambda=\lambda_c\approx 0.6$ (Fig.8). In the case of uniform friction the 
critical potential-asymmetry $b\approx 0.25$ which is what has been used 
incidentally by Mateos\cite{Mateos1} and others corresponding to their parameter values.
 As the value of $\lambda$  is changed, the nature of the particle trajectories changes
 qualitatively.

\begin{figure}[t]
  \centering
  \subfigure[]{\includegraphics[width=12cm,height=6.5cm,angle=-90]{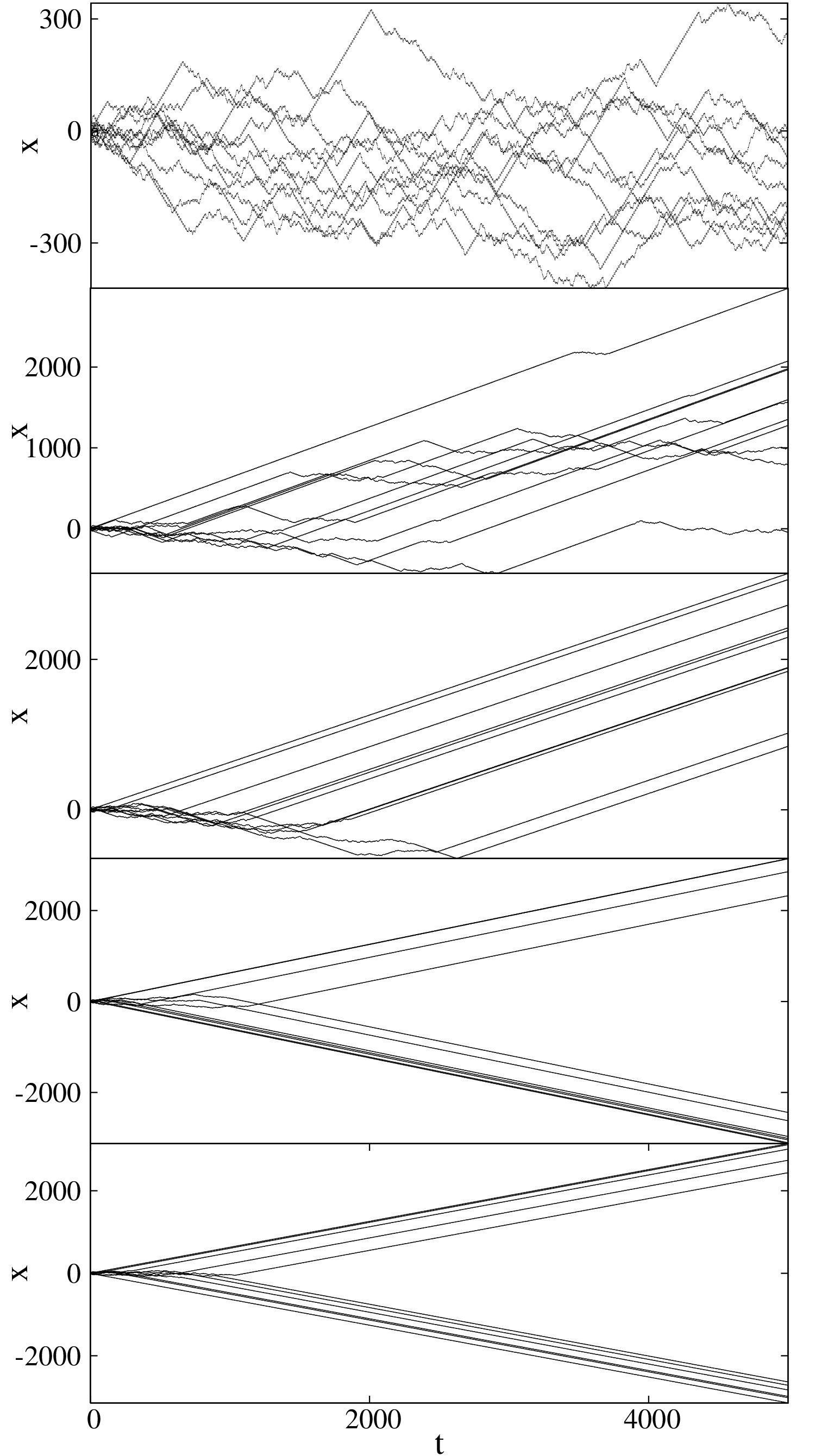}}
\hspace{0.4cm}
\subfigure[]{\label{fig:edge-c}\includegraphics[width=12cm,height=6.5cm,angle=-90]{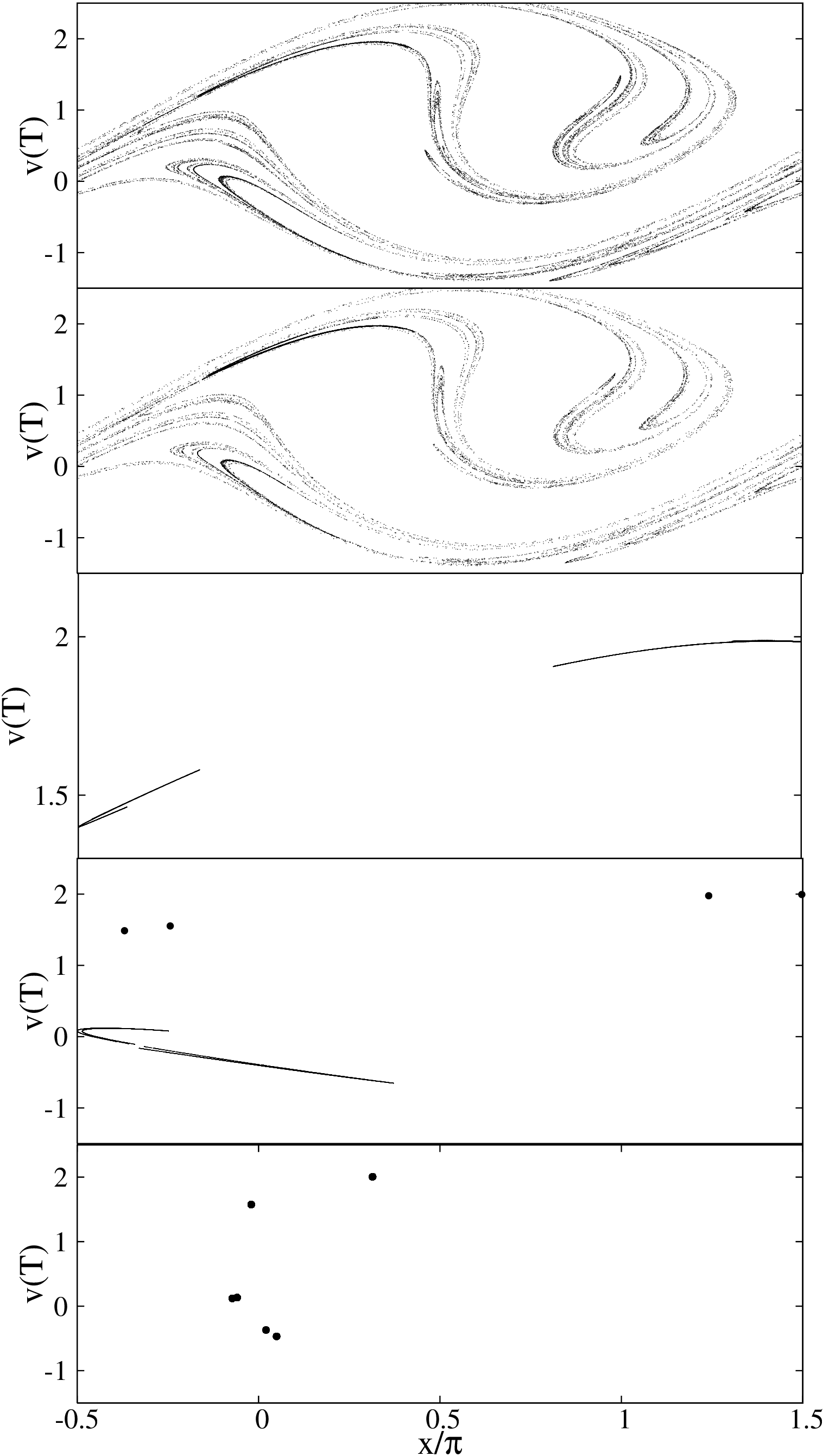}}
\caption{Particle trajectories (a), and the corresponding stroboscopic plots (b), for 
different values of inhomogeneity parameter $\lambda$ ($\lambda = 0.9$ (top) to 
$\lambda = 0.5$ (bottom). $\tau = 10$, $a = 0.67$, $\gamma = 0.12$, $\phi = 0.35$.
Trajectories are shown for ten different sample initial conditions. Though the trajectories
appear to be straight line in the scale of the figures, they are oscillatory as in Fig.3}
\label{fig:edge}
\end{figure}

In Fig.9 we present the typical trajectories $x(t)$ along with the stroboscopic
phase points at regular time intervals $t=n\tau,n=1,2,3, ..$ for various values 
of the inhomogeneity parameter $\lambda$ for amplitude $a=0.67$, period
$\tau=10$ and initial phase $\phi_0=0$ of the external field $F(t)$. For 
$\lambda=0.9$ all the trajectories are chaotic but nontransporting giving zero 
ratchet current. For $\lambda=0.8$ the trajectories are chaotic consisting of
longer quasiperiodic transporting trajectories in the positive direction but
interrupted by shorter similar trajectories in the negative direction giving
finite ratchet current in the positive direction. For $\lambda=0.7$ the current
$\langle\langle v \rangle\rangle$ is close to the maximum (=$2\pi/\tau$) 
corresponding to the fundamental velocity. In this case all the trajectories 
are quasiperiodic transporting in the same positive direction. The 
corresponding stroboscopic plot shows their chaotic nature. As the $\lambda$
is lowered from 0.7 to 0.6 we obtain a mixture of transporting quasiperiodic 
trajectories and transporting regular periodic trajectories. However, as the 
value $\lambda=0.6$ is reached their directions of average motion gradually
become equal giving zero current at $\lambda=0.6$. As mentioned earlier,
$\langle\langle v \rangle\rangle$ is independent of $\phi_0$ for the 
considered parameters. For values of $\lambda\leq 0.5$ all the 
trajectories are transporting regular periodic with same average slope but
differing in direction. For $\phi_0=0$ there are more negative-slope 
trajectories and vice-versa for $\phi_0=\pi$. The five-paneled Fig.9 gives
a qualitative mechanism for the explanation of Fig.8 and underlines the 
importance of averaging over the initial phase $\phi_0$ at smaller 
inhomogeneities.

\subsection{Role of noise and low amplitudes of drives}
  In deterministic ratchets, because of the complex structure of 
the basins of attraction and the presence of multiple attractors in phase
space\cite{Mateos2}, the nature of the current is sensitively dependent
on the initial conditions.  The presence of noise in the system, however, 
makes the particle dynamics independent of the initial conditions allowing
 the particle to explore all the existing attractors. As a result the 
chaotic behaviour of the system disappears.
\begin{figure}
\begin{center}
\epsfig{file=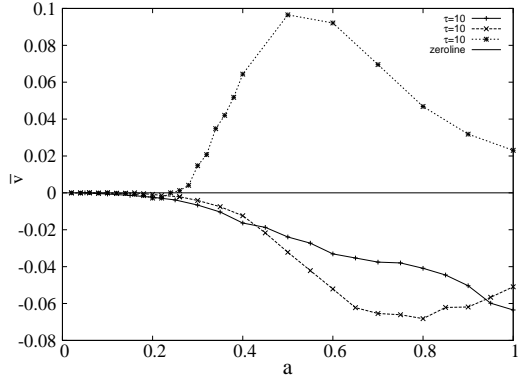, angle=-90, width=7.0cm,totalheight=5cm}
\caption{Plot of $\bar v $ versus $a$ in the presence of noise, for 
three values of $\tau$; $T=0.4$, $\gamma = 0.12$, $\lambda = 0.9$, $\phi = 0.35$.}
\end{center}
\end{figure}
The average velocity shows a relatively smoother variation with the  amplitude
of external drive than in the deterministic case. In Fig.10, the variation of 
$\bar v$ with $a$ is shown for three different values of drive frequencies, 
keeping all the other parameters fixed.
\par With noise, appreciable ratchet current is obtained even at low amplitudes, 
where there was no current in the deterministic case (Fig.10). This is because, 
the presence of noise aids the particles to overcome the potential barrier which 
it could not have done otherwise. Hence the presence of noise plays a 
non-negligible role in obtaining ratchet current in the low amplitude regime, 
contrary to what has been mentioned by others \cite{Mateos1}. The same effect of
noise has been shown in a different system using the matrix continued fraction 
method\cite{Denisov}.

\section{Conclusion}
 In this work the deterministic ratchet current was obtained in a periodically 
driven symmetric potential in the presence of frictional inhomogeniety (most
of the earlier works were with a driven asymmetric potential). The particle 
dynamics shows similar characteristic as in an asymmetric potential. Initial 
conditions and the control parameters play a major role in shaping the nature 
of the dynamics. For low amplitudes of drive the initial phase of the drive is 
found to play a non-negligible role. At these amplitudes, we show that averaging 
over initial conditions and initial phases of the external drive is essential to 
obtain realistic averages. With noise, ratchet current is obtained at amplitude 
regimes where there is no current in the deterministic systems.

\begin{center}
{\Large{\it{Acknowledgement}}}
\end{center}
MCM acknowledges partial finacial support from BRNS, DAE, Govt. of India under 
the grant No.2009/37/17/BRNS.

\end{document}